\documentclass[12pt]{article}
\usepackage{bezier}
\usepackage{amssymb}
\usepackage{epsfig}

\newcommand{\nc}{\newcommand}
\nc{\be}{\begin{equation}}
\nc{\ee}{\end{equation}}
\nc{\bea}{\begin{eqnarray}}
\nc{\eea}{\end{eqnarray}}

\nc{\eqn}[1]{{(\ref{#1})}}
\nc{\cA}{{\cal A}}
\nc{\cB}{{\cal B}}
\nc{\cC}{{\cal C}}
\nc{\cD}{{\cal D}}
\nc{\cE}{{\cal E}}
\nc{\cF}{{\cal F}}
\nc{\cG}{{\cal G}}
\nc{\cH}{{\cal H}}
\nc{\cI}{{\cal I}}
\nc{\cJ}{{\cal J}}
\nc{\cK}{{\cal K}}
\nc{\cL}{{\cal L}}
\nc{\cM}{{\cal M}}
\nc{\cN}{{\cal N}}
\nc{\cO}{{\cal O}}
\nc{\cP}{{\cal P}}
\nc{\cQ}{{\cal Q}}
\nc{\cR}{{\cal R}}
\nc{\cS}{{\cal S}}
\nc{\cT}{{\cal T}}
\nc{\cU}{{\cal U}}
\nc{\cV}{{\cal V}}
\nc{\cW}{{\cal W}}
\nc{\cX}{{\cal X}}
\nc{\cY}{{\cal Y}}
\nc{\cZ}{{\cal Z}}
\nc{\simo}[1]{{\stackrel{#1}{\simeq}}}
\nc{\geqo}[1]{{\stackrel{#1}{\geq}}}
\nc{\geo}[1]{{\stackrel{#1}{>}}}
\nc{\guo}[1]{{\stackrel{#1}{\succ}}}

\nc{\rbo}{\raisebox}
\nc{\RR} {\rangle \! \rangle}
\nc{\LL} {\langle \! \langle}
\nc{\rmi}[1]{{\mbox{\small #1}}}
\nc{\eq}{eq.~}
\nc{\nr}[1]{(\ref{#1})}
\nc{\ul}{\underline}
\nc{\mc}{\multicolumn}
\nc{\todo}[1]{\par\noindent{\bf $\rightarrow$ #1}}

\nc{\cu}{{\cal u}}

\title{
  \begin{flushright} {\small $\begin{array}{ l } \mbox{HD--THEP--99--08} \\
    \end{array} $}
 \end{flushright}
The axial anomaly in lattice QED. \\
A universal point of view}

\author{T.~Reisz\thanks{Supported by a Heisenberg Fellowship}
        $\;$ and
       H.~J.~Rothe,
         \\ \\Institut
        f\"ur Theoretische Physik,\\
        Universit\"at Heidelberg, \\
        Philosophenweg 16, \\
        D-69120 Heidelberg, Germany}

\begin{document}

\maketitle

\begin{abstract}
We give a perturbative proof that U(1) lattice gauge theories
generate the axial anomaly in the continuum limit
under very general conditions on the lattice Dirac operator.
These conditions are locality, gauge covariance and the absense 
of species doubling.
They hold for Wilson fermions as well as for realizations of 
the Dirac operator that satisfy the Ginsparg-Wilson
relation.
The proof is based on the lattice power counting theorem.
\end{abstract}

%
%
%
%
The axial anomaly in
lattice QED with Wilson fermions has first been studied
by Karsten and Smit \cite{karsten}. 
They showed that in the continuum limit
it arises from an irrelevant operator in the lattice Ward identity
originating from the Wilson term in the action.
That this operator must necessarily generate the ABJ anomaly \cite{adler}
in the continuum limit has been shown in \cite{rothe} using the
small $a$ expansion scheme of \cite{wetzel}.
Recently, an alternative lattice regularization of the Dirac operator
has been discovered which posseses an exact chiral symmetry on
the lattice and is free of fermion species doubling and other 
undesired lattice artifacts
\cite{neuberger_luescher}.
The no-go theorem of Nielsen and Ninomiya \cite{nielsen_ninomiya} is 
circumvented by choosing the Dirac operator to satisfy the
Ginsparg-Wilson relation \cite{ginsparg_wilson}.
Within this context, the ABJ anomaly on the lattice has been studied in
\cite{luescher_chiu_hasenfratz}.

In the continuum formulation of QED one knows that various
regularizations that preserve gauge invariance all yields the
same expression for the anomaly when the cutoff is removed.
One therefore should expect that the anomaly will also be reproduced in
the continuum limit for any lattice regularization of the 
Dirac operator that satisfies a general set of conditions.
The purpose of this letter is to show that this is indeed the case.

Consider a fermionic action of the form
\be \label{ano.ferm}
  S_{ferm} \; = \; a^4 \sum_{x\in a {\mathbb Z}^4} \overline{\psi}(x)
   \biggl( \left( D[U]+m \right) \psi \biggr)(x) ,
\ee
with $a$ the lattice spacing, $m\ne 0$ the fermion mass,
and $D[U]$ a lattice regularization of the Dirac operator.
$U=\{U(x;\mu)\}$ represent the set of $U(1)$ link variables.
Summation in (\ref{ano.ferm}) is over all lattice sites.
We parametrize the link variables $U(x;\mu)$ according to
\be \label{ano.param}
 U(x;\mu)=\exp{(ia A_\mu(x))},
\ee
with real valued gauge potential $A_\mu(x)$.
An observable $\cO(\psi,\overline{\psi},A)$ is gauge invariant
if
\be
  \cO(\psi^\omega ,\overline{\psi}^{\,\omega}, A^\omega ) \; = \;
  \cO(\psi ,\overline{\psi}, A),
\ee
with $\omega(x)\in{\mathbb R}$, where
\bea \label{ano.gauge}
  && \psi^{\omega}(x) = e^{i\omega(x)} \psi(x),\;
  \overline{\psi}^{\,\omega}(x) = e^{-i\omega(x)} \overline{\psi}(x),
  \nonumber \\
  && A_\mu^{\omega}(x) = A_\mu(x) 
   - \frac{1}{a}\widehat\partial_\mu\omega(x).
\eea
We denote the forward and backward lattice differences by
\be
  \widehat\partial_\mu f(x) = f(x+a\widehat\mu) - f(x),\;
   \widehat\partial_\mu^* f(x) = f(x) - f(x-a\widehat\mu).
\ee
The action is assumed to satisfy the following conditions:
\begin{enumerate}
\item{\it Gauge invariance}.
Under a gauge transformation (\ref{ano.gauge}) $S_{ferm}$
is invariant,
\be
  S_{ferm}(\psi^\omega ,\overline{\psi}^{\,\omega}, A^\omega ) \; = \;
  S_{ferm}(\psi ,\overline{\psi}, A).
\ee
\item{\it Continuum limit}.
The action $S_{ferm}$ converges for $a\to 0$ 
to the standard continuum fermion action,
\be
  S_{ferm} \; \to \; \int d^4x \;
   \overline\psi(x) \left( \sum_{\mu=0}^3 \gamma_\mu 
    ( \partial_\mu + i A_\mu ) + m \right) \psi(x).
\ee

\item{\it Locality}.
$D[\exp{iaA}]$ allows for a formal small field expansion
\bea \label{ano.small}
  && D[\exp{iaA}](x,y) \; = \; \sum_{n\geq 0} \frac{1}{n!}
   a^{4n} \sum_{z_1,\dots ,z_n} \sum_{\mu_1,\dots ,\mu_n}
   D^{(n)}_{\mu_1\dots\mu_n}(x,y \vert z_1,\dots ,z_n)
  \nonumber \\
  && \qquad\qquad\qquad\qquad\qquad \cdot  
   A_{\mu_1}(z_1) \cdots A_{\mu_n}(z_n),
\eea
where the coefficient functions $D^{(n)}_{\nu_1\cdots\nu_n}$ are 
local in the sense that
they fall off exponentially
fast for large separations between any pair of sites, with a decay constant 
proportional to the inverse lattice spacing.

\item{\it $D^{(0)}$ is free of doublers}.
The Fourier transform of the free Dirac
propagator, ${\widetilde{D}}^{0}$, is invertible for non-vanishing momentum 
(modulo the Brillouine zone $2\pi/a$).
\end{enumerate}

Condition 3.~implies that the Fourier transforms $\widetilde{D}^{(n)}$
of $D^{(n)}$ are analytic functions around  zero momentum.
This property also holds for Ginsparg-Wilson fer\-mions.
It is a consequence of the locality of the Dirac operator
$D(U)$, which for sufficiently smooth gauge fields has been shown 
to hold in \cite{luescher_jansen}.

Let us define the expectation value of an observable 
$\cO(\psi,\overline\psi,A)$ in an external gauge field $A$ by
\be
  < \cO >_A \; = \; \frac{1}{Z(A)} \;
   \int \prod_x d\psi(x) d\overline\psi(x) \; \cO(\psi,\overline\psi,A)
  e^{-S_{ferm}(\psi,\overline\psi ,A)},
\ee
with $Z(A)$ such that $<1>_A=1$.
Then under the above conditions the following statement holds.
There exists a gauge invariant axial vector current 
$j_\mu^5(x)$ for which
\be \label{ano.div}
  \lim_{m\to 0} \lim_{a\to 0} \sum_{\mu=0}^3 \frac{1}{a}
   \widehat\partial_\mu^* < j_\mu^5(x) >_A \; = \;
   \frac{1}{16\pi^2} \sum_{\mu\nu\rho\lambda}
  \; \epsilon_{\mu\nu\rho\lambda}
   F_{\mu\nu}(x) F_{\rho\lambda}(x),
\ee
where $F_{\mu\nu}(x) = \partial_\mu A_\nu(x) - \partial_\nu A_\mu(x)$.
$j_\mu^5$ is local in the same sense as $D[\exp{iA}]$ is according to
assumption 3, with
classical continuum limit
\be \label{ano.classical}
 \lim_{a\to 0} j_\mu^5(x) \; = \;
  \overline\psi(x)\gamma_\mu\gamma_5\psi(x).
\ee
The proof of (\ref{ano.div}) proceeds in three steps.

i)
Consider the standard infinitessimal axial transformation
\bea \label{ano.transf}
  \delta\psi(x) & = & i \omega(x) \gamma_5 \psi(x) , 
   \nonumber \\
  \delta\overline\psi(x) & = & i \omega(x) \overline\psi(x) \gamma_5 .
\eea
Since the fermion measure is invariant under this transformation
we have
\be
  0 \; = \; \delta <\cO >_A \; = \;
   < - \cO \delta S + \delta\cO >_A .
\ee
Because of the assumptions 2.~and 3.~the variation of the action can
always be written in the form
\be \label{ano.deltas}
  \delta S \; = \; i a^4 \sum_{x\in\mathbb{Z}^4} \omega(x) 
   \left\lbrack - \frac{1}{a} \sum_\mu \widehat\partial_\mu^* j_\mu^5(x)
   + Q^5(x) \right\rbrack,
\ee
with $j_\mu^5$ a local dimension 3 operator satisfying 
(\ref{ano.classical}).
According to condition 1, $j_\mu^5$ can always be chosen to be gauge
invariant.
Furthermore, $Q^5$ has the form
\be \label{ano.qdecompose}
  Q^5(x) \; = \; 2 m \overline\psi(x) \gamma_5 \psi(x) \; + \;
   \Delta(x) ,
\ee
with $\Delta$ a local gauge invariant irrelevant operator
(that is, it vanishes in the classical continuum limit)
of canonical dimension 4.
Writing 
\be
  \delta\cO \; = \; i a^4 \sum_{x} \omega(x) 
   \left(\cT\cO\right)(x),
\ee
we obtain the local axial vector Ward identity
\be \label{ano.cwi_base}
  < \frac{1}{a} \widehat\partial_\mu^* j_\mu^5(x) \, \cO >_A \; = \;
   < Q^5(x) \cO - \left( \cT\cO\right)(x) >_A.
\ee
Except for the properties listed above, 
the precise structure of the 
lattice current $j_\mu^5$ and of $Q^5$ will not be needed in the following.

At this point we remark that in the case where the Dirac operator is 
chosen to satisfy the Ginsparg-Wilson relation, the action possesses 
(for $m=0$) an exact 
chiral symmetry generated by e.g.
\bea
  \delta\psi(x) & = & i \omega(x) \gamma_5 
   \left( (1 - \frac{1}{2} aD)\psi \right)(x) , 
   \nonumber \\
  \delta\overline\psi(x) & = & i \omega(x) 
   \left(\overline\psi (1-\frac{1}{2} aD) \right)(x) \gamma_5 .
\eea
$aD$ is an irrelevant lattice operator
of dimension zero. The fermionic measure is however not invariant 
under this transformation. Taking account of this, the Ward identity
can be cast again into the form (\ref{ano.cwi_base}).

ii)
Setting $\cO = 1$ in (\ref{ano.cwi_base}) we have 
\be \label{ano.cwi}
  < \frac{1}{a} \sum_\mu \widehat\partial_\mu^* j_\mu^5(x) >_A \; = \;
   < Q^5(x) >_A ,
\ee
with $j_\mu^5$ a local, gauge invariant dimension 3 operator with
classical continuum limit $\overline\psi(x)\gamma_\mu\gamma_5\psi(x)$,
and $Q^5$ given by (\ref{ano.qdecompose}) ,with $\Delta$ an
irrelevant gauge invariant operator of dimension 4.

In the following we work in momentum space for convenience.
Let us write
\be
   A_\mu(x) = \int_k  
   \widetilde{A}_\mu(k) \; e^{i k\cdot x + i k_\mu a/2} , \;
   \int_k \equiv \int_{-\pi/a}^{\pi/a} \frac{d^4k}{(2\pi)^4} ,
\ee
and
\bea
  &&  < j_\mu^5(x) >_A  =  \int_q 
   \widetilde{j}_\mu^5(q) \; e^{i q\cdot x +i q_\mu a/2} , \;
   < Q^5(x) >  =  \int_q 
   \widetilde{Q}^5(q) \; e^{i q\cdot x} ,
  \nonumber \\
  && \widetilde{j}_\mu^5(q) = \sum_{n\geq 2} \frac{1}{n!}
   \sum_{\nu_1,\dots ,\nu_n}
   \int_{p_1,\dots ,p_{n-1}} 
   \widehat{\Gamma}^{5,\mu}_{\nu_1\dots\nu_n}(q\vert p_1,\dots ,p_{n-1})
  \nonumber
  \\
  && \qquad\qquad \cdot \widetilde{A}_{\nu_1}(p_1) \dots
   \widetilde{A}_{\nu_n}(q-p_1-\dots -p_{n-1}),
  \\
  && \widetilde{Q}^5(q) = \sum_{n\geq 2} \frac{1}{n!}
   \sum_{\nu_1,\dots ,\nu_n}
   \int_{p_1,\dots ,p_{n-1}} 
   \widehat{\Gamma}^{5}_{\nu_1\dots\nu_n}(q\vert p_1,\dots ,p_{n-1})
  \nonumber
  \\
  && \qquad\qquad \cdot \widetilde{A}_{\nu_1}(p_1) \dots
   \widetilde{A}_{\nu_n}(q-p_1-\dots -p_{n-1}) .
  \nonumber
\eea
Here and in the following the explicit $a$-dependence of the
correlation functions $\widehat{\Gamma}$ is suppressed.
The chiral Ward identity (\ref{ano.cwi}) then becomes
\be \label{ano.cwi_pspace}
  i \sum_{\mu=0}^3 \widehat{q}_\mu 
   \widehat{\Gamma}^{5,\mu}_{\nu_1\cdots\nu_n}(q\vert p_1,\dots ,p_{n-1})
   \; = \;
   \widehat{\Gamma}^{5}_{\nu_1\cdots\nu_n}(q\vert p_1,\dots ,p_{n-1}),
\ee
where $\widehat{q}_\mu = (2/a)\sin{(q_\mu a/2)}$.
Beause the fermion measure is invariant under
(\ref{ano.gauge}), every gauge invariant quantity $\cO$
satisfies
\be
  < \cO(\psi,\overline\psi, A^\omega) >_{A^\omega} \; = \;
  < \cO(\psi,\overline\psi, A) >_{A}.
\ee
This implies the gauge Ward identity
\be
  0 \; = \; \sum_\nu \frac{1}{a} \widehat\partial_{z_\nu}^* 
   \frac{\partial}{\partial A_\nu(z)}
   < \cO(\psi,\overline\psi ,A) >_A.
\ee
Since the fermion mass is different from zero, and by assumption 3,
the correlation functions $\Gamma^{5,\mu}$ are analytic around
zero momenta.
Writing $\cT^\delta$ for the Taylor operation of order $\delta$
applied around zero momentum,
one finds that gauge invariance implies
\be \label{ano.taylor}
  \cT^1_{p,q} \widehat{\Gamma}^{5,\mu}_{\nu_1\dots\nu_n}
   (q\vert p_1,\dots ,p_{n-1}) \; = \; 0 .
\ee

iii)
Finally consider the contribution to the external field expectation value 
of the divergence of the axial vector current involving two external 
photon lines. It is this contribution which is expected to generate the 
anomaly. 
Using (\ref{ano.taylor}) we write
\bea \label{ano.divsubtr}
  && i \sum_\mu \widehat{q}_\mu \widehat{\Gamma}^{5,\mu}_{\nu_1\nu_2}(q\vert p)
   \; = \; i \sum_\mu \widehat{q}_\mu \left( 1 - \cT^1_{q,p} \right) 
   \widehat{\Gamma}^{5,\mu}_{\nu_1\nu_2}(q\vert p) 
  \nonumber \\
  && = i \left( 1 - \cT^2_{q,p} \right) \sum_\mu \widehat{q}_\mu
   \widehat{\Gamma}^{5,\mu}_{\nu_1\nu_2}(q\vert p) 
   \; = \; \left( 1 - \cT^2_{q,p} \right) 
   \widehat{\Gamma}^5_{\nu_1\nu_2}(q\vert p) ,
\eea
where in the last step we have used the chiral Ward identity
(\ref{ano.cwi_pspace}).
The absense of species doubling (assumption 4) allows one to obtain
the continuum limit by applying the lattice power counting theorem
\cite{reisz}.
Because $Q^5$ has canonical dimension 4,
the lattice divergence degree of every Taylor-subtracted
lattice Feynman integral that contributes to the RHS of
(\ref{ano.divsubtr}) has negative lattice divergence degree.
Writing generically
\be
 \widehat \Gamma  = \; \int_{-\pi/a}^{\pi/a} \frac{d^4k}{(2\pi)^4} \;
   \widehat I,
\ee
the continuum limit is thus obtained by taking the limit under
the integral sign and sending the integration limits to 
infinity. All contributions of the irrelvant parts of $\widehat{\Gamma}^5$
that originate from $\Delta$ vanish.
With the notation
$I \; = \; \lim_{a\to 0} \widehat{I}$
we get
\bea \label{ano.atozero}
 && \lim_{a\to 0} \left( 1 - \cT^2_{q,p} \right)  
\widehat{\Gamma}^5_{\nu_1\nu_2}(q\vert p)
   = \int_{-\infty}^\infty \frac{d^4k}{(2\pi)^4} \;
   \lim_{a\to 0} \left( 1 - \cT^2_{q,p} \right) 
    \widehat{I}^5_{\nu_1\nu_2}(q,p,k) 
  \nonumber \\
= &&\int_{-\infty}^\infty \frac{d^4k}{(2\pi)^4} \;
   \left( 1 - \cT^2_{q,p} \right) 
    I^{5,m}_{\nu_1\nu_2}(q,p,k) ,
\eea
which is the continuum Feynman integral for the triangle graphs with a 
$2m\gamma_5$ insertion, Taylor subracted at zero momenta to second order. 
Since these triangle graphs have an ultraviolet degree of divergence
1, we can further write
\bea
 && \lim_{a\to 0} \left( 1 - \cT^2_{q,p} \right)  
 \widehat{\Gamma}^5_{\nu_1\nu_2}(q\vert p)
 \nonumber \\
  && = - \int_{-\infty}^\infty \frac{d^4k}{(2\pi)^4} \;
    \left( \cT^2_{q,p} - \cT^1_{q,p} \right) I_{\nu_1\nu_2}^{5,m}(q,p,k) 
    + \int_{-\infty}^\infty \frac{d^4k}{(2\pi)^4} \;
    \left( 1 - \cT^1_{q,p} \right) I_{\nu_1\nu_2}^{5,m}(q,p,k) 
 \nonumber \\
 && = - \int_{-\infty}^\infty \frac{d^4k}{(2\pi)^4} \;
    \left( \cT^2_{q,p} - \cT^1_{q,p} \right) I_{\nu_1\nu_2}^{5,m}(q,p,k) 
    + O(m \; \log{m}),
\eea
where the last step holds for non-exceptional momenta.
Finally, we let $m\to 0$, with the result
\bea \label{ano.chiral}
  && \lim_{m\to 0} \lim_{a\to 0} i \sum_\mu \widehat{q}_\mu 
   \widehat{\Gamma}^{5,\mu}_{\nu_1\nu_2}(q\vert p)
  = - \lim_{m\to 0}
    \int_{-\infty}^\infty \frac{d^4k}{(2\pi)^4} \;
     \left( \cT^2_{q,p} - \cT^1_{q,p} \right) 
   I_{\nu_1\nu_2}^{5,m}(q,p,k) 
  \nonumber \\
 && = - \lim_{m\to 0} 16 m^2 \int_{-\infty}^\infty 
   \frac{d^4k}{(2\pi)^4} \; \frac{1}{(k^2+m^2)^3}
  \; \sum_{\alpha \beta} 
   \epsilon_{\nu_1\nu_2\alpha\beta} q_\alpha p_\beta \\
 && = - \frac{1}{2\pi^2} \sum_{\alpha \beta} 
   \epsilon_{\nu_1\nu_2\alpha\beta} q_\alpha p_\beta ,
 \nonumber
\eea
which is the axial anomaly.

Finally, we show that higher order corrections involving more than two 
external photon lines vanish in the chiral limit. Consider the chiral 
Ward identity (22) for $n\geq 3$,
\bea
  && i \sum_\mu \widehat{q}_\mu 
   \widehat{\Gamma}^{5,\mu}_{\nu_1\dots\nu_n}
   (q\vert p_1,\dots ,p_{n-1}) \; = \;    
  \nonumber \\
  && = \widehat\Gamma^{5,m}_{\nu_1\dots\nu_n}
   (q\vert p_1,\dots ,p_{n-1}) \; + \;
   \widehat\Gamma^{5,\Delta}_{\nu_1\dots\nu_n}(q\vert p_1,\dots ,p_{n-1}) 
  \nonumber \\
  && = \left( 1-\cT^{3-n}_{q,p} \right)
   \widehat\Gamma^{5,m}_{\nu_1\dots\nu_n}(q\vert p_1,\dots ,p_{n-1}) 
  \\ 
  && + \left( 1-\cT^{4-n}_{q,p} \right)
   \widehat\Gamma^{5,\Delta}_{\nu_1\dots\nu_n}(q\vert p_1,\dots ,p_{n-1}),
  \nonumber \\
\eea
where $\cT^\delta_{q,p}=0$ if $\delta<0$.
$\widehat\Gamma^{5,m}$ and $\widehat\Gamma^{5,\Delta}$ denote the 
contribution to $\widehat\Gamma^5$ of
$2m\overline\psi\gamma_5\psi$ and $\Delta$, respectively.
We have used the gauge invariance of
$\Gamma^{5,m}$ and $\Gamma^{5,\Delta}$,
which implies that both observables
satisfy a relation analogous to 
(\ref{ano.taylor}).
Again by the lattice power counting theorem,
the second term of the last equality above vanishes as $a\to 0$, 
whereas the first term has a finite continuum limit.
This term vanishes for non-exceptional momenta,
upon taking the chiral limit $m\to 0$.
This completes the proof.

We conclude with a remark.
The irrelevant lattice operator $\Delta$ (cf.~(\ref{ano.qdecompose})),
of course plays an important role for generating the anomaly.
Indeed, the anomaly can be represented as the continuum limit
of an irrelevant lattice Feynman integral originating from $\Delta$.
For, applying $(\cT^2_{q,p}-\cT^1_{q,p})$ to 
both sides of the Ward identity (\ref{ano.cwi_pspace}) with $n=2$,
and making use of gauge invariance, 
the left hand side vanishes, and we obtain
\be
 \cT^2_{p,q}
   \widehat{\Gamma}^{5,\Delta}_{\nu_1\nu_2}(q\vert p) 
  = \left( \cT^2_{p,q} - \cT^1_{p,q} \right)
   \widehat{\Gamma}^{5,\Delta}_{\nu_1\nu_2}(q\vert p)
  = - \left( \cT^2_{p,q} - \cT^1_{p,q} \right)
   \widehat{\Gamma}^{5,m}_{\nu_1\nu_2}(q\vert p) .
\ee
The continuum limit of the right hand side is computed similar as before,
by applying the lattice power counting theorem,
and yields the axial anomaly, 
\be
 \lim_{m\to 0} \lim_{a\to 0}
   \widehat{\Gamma}^{5,\Delta}_{\nu_1\nu_2}(q\vert p)
 = \lim_{m\to 0} \lim_{a\to 0}
  \left( 1 - ( 1 - \cT^2_{p,q} ) \right)
   \widehat{\Gamma}^{5,\Delta}_{\nu_1\nu_2}(q\vert p)
 = - \frac{1}{2\pi^2} \sum_{\alpha \beta} 
   \epsilon_{\nu_1\nu_2\alpha\beta} q_\alpha p_\beta ,
\ee
irrespective of the particular form of $\Delta$.

%
%




\begin{thebibliography}{9}
\bibitem{karsten} L.~H.~Karsten and J.~Smit,
Nucl.~Phys.~{\bf B183} (1981) 103.
\bibitem{adler} S.~Adler, Phys.~Rev.~{\bf 177} (1969) 2426;
J.~S.~Bell and R.~Jackiw, Nuovo Cimento~{\bf 60A} (1969) 47;
S.~Adler and W.~A.~Bardeen, Phys.~Rev.~{\bf 182} (1969) 1517.
\bibitem{rothe} H.~J.~Rothe and N.~Sadooghi,
Phys.~Rev.~{\bf D58} (1998) 074502.
\bibitem{wetzel} W.~Wetzel, Nucl.~Phys.~{\bf B255} (1985) 659
\bibitem{neuberger_luescher} 
H.~Neuberger, Phys.~Lett.~{\bf B417} (1998) 141; 
{\sl ibid} {\bf B427} (1998) 353;
M.~L\"uscher, Phys.~Lett.~{\bf B428} (1998) 342
\bibitem{nielsen_ninomiya} H.~B.~Nielsen and M.~Ninomiya, 
Nucl.~Phys.~{\bf B185} (1981) 20; ibid {\bf B193} (1981) 173.
\bibitem{ginsparg_wilson} P.~Ginsparg and K.~Wilson,
Phys.~Rev.~{\bf D25} (1982) 2649
\bibitem{luescher_chiu_hasenfratz} 
P.~Hasenfratz, V.~Laliena and F.~Niedermayer,
Phys.~Lett.~{\bf B427} (1998) 125;
M.~L\"uscher, Nucl.~Phys.~{\bf B538} (1999) 515;
Ting-Wai Chiu, Phys.~Lett.~{\bf B445} (1999) 371;
Ting-Wai Chiu and Tung-Han Hsieh, e-Print Archive hep-\-lat\-/9901011.
\bibitem{luescher_jansen}
P.~Hern$\acute{a}$ndez, K.~Jansen, M.~L\"uscher, 
e-Print Archive hep-\-lat\-/9808010.
\bibitem{reisz} T.~Reisz, Comm.~Math.~Phys.~{\bf 116} (1988) 81;
M.~L\"uscher, ``Selected Topics in Lattice Field Theory'',
Lectures given at the Summer school on {\sl Fields, Strings
and Critical Phenomena}, Les Houches (1988);
H.~J.~Rothe, ``Lattice Gauge Theories: An Introduction'',
Second Edition, World Scientific Lecture Notes in Physics Vol.~59,
World Scientific 1997.

\end{thebibliography}
\end{document}